\documentclass[a4paper]{jpconf}
\usepackage{graphicx}
\usepackage{amsmath,amsfonts,amssymb,amsthm}
\usepackage[amssymb]{SIunits}
\usepackage{graphicx}
\usepackage{caption}
\usepackage{color}
 \usepackage{subfig}

\begin{document}
\title{Search for $\text{H}\to \text{b}\bar{\text{b}}$ in association with single top quarks as a test of Higgs couplings}

\author{Benedikt Maier on behalf of the CMS collaboration}

\address{Institut f\"ur Experimentelle Kernphysik, Karlsruhe, Germany}

\ead{benedikt.maier@cern.ch}

\begin{abstract}
The associated production of Higgs boson and single top quark is of particular interest since it is senstive to the relative sign of the Higgs boson coupling to gauge bosons and the Yukawa coupling $y$ to fermions. The presented analysis is setting upper production limits on a model with $y_\text{t}=-1$, which has an enhanced cross section compared to the standard model expectation. For this it focusses on the Higgs boson decaying to a pair of b quarks and uses the full dataset of $pp$ collisions recorded with the CMS detector in 2012. It reports an upper limit on 7.57 times the expected cross section, with an expected sensitivity of 5.14. This translates into the exclusion of associated tHq production with $y_\text{t}=-1$-like characteristics with a cross section smaller than 1.77\,pb.
\end{abstract}

\section{Introduction}
Since the discovery of a Higgs boson-like particle in 2012 \cite{ATLAS,CMS}, the ATLAS \cite{ATLAS_name} and CMS collaborations \cite{CMS_name} are probing its characteristics in order to test whether it is exactly the Higgs boson as predicted by the standard model (SM). One of its important properties is the Yukawa coupling mechanism to fermions, where coupling strengths are proportional to the fermion masses. In particular, the coupling to the top quark, $y_\text{t}$, is a significant parameter for verification of the electroweak sector of the SM, since it is the heaviest particle known to exist. While the absolute value of $y_\text{t}$ is estimated to be close to 1 in recent measurements \cite{ATLAS2,CMS2} and hence in accordance with what the SM predicts, most channels are insensitive to its sign, and the anomalous scenario of $y_\text{t}=-1$ is not ruled out yet by either experiment. 

\begin{figure}[htpb]
\centering
\includegraphics[scale=1.2]{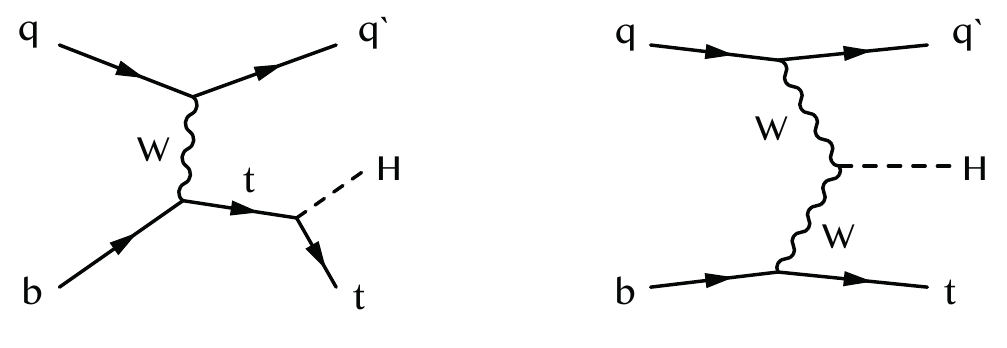}
\caption{Leading order Feyman diagrams for the $t$-channel single top quark production in association with a Higgs boson}
\label{fig:feynman}
\end{figure}

The associated production of Higgs boson and single top quark is highly sensitive to the value of $y_\text{t}$ and to the relative sign of the Higgs boson coupling to the top quark and to vector bosons. For the standard model, there exists a destructive interference between the two leading order processes of associated production of single top quark and Higgs boson shown in Figure~\ref{fig:feynman}, resulting in a cross section of $\sigma_{\text{tHq}}=18.3$\,fb \cite{farina}. Changing the coupling strength $y_\text{t}$ gives a cross section enhanced by a factor of $\sim 13$ for $y_\text{t}=-1$, which brings this scenario close to the reach for searches with the integrated luminosity collected at $\sqrt{s}=8$\,TeV.

\section{Analysis overview}

\subsection{Event selection}

This analysis is optimized for the $y_\text{t}=-1$ scenario, which has a production cross section of $\sigma_{\text{tHq}}=234$\,fb, and focusses on $\text{H}\to \text{b}\bar{\text{b}}$ decays. Together with the b~quark from the leptonically decaying top quark and the light flavour jet typical for $t$-channel single top events, this gives a unique signature with high jet and b jet multiplicities, where an additional b jet is expected to come from the gluon splitting which is needed to initiate the $t$-channel single top production.

An event in the muon (electron) channel needs to have exactly one muon (electron) with a transverse momentum $p_\text{T}>26\,(30)$\,GeV and a pseudorapidity $|\eta|<2.1\,(2.4)$. These criteria in connection with requirements related to lepton isolation define a tight muon (electron). Additional loose leptons with relaxed cuts are vetoed in each event, leading to the rejection of Drell-Yan + jets processes.

The final state of a signal event in this channel contains three or four b~jets as well as one untagged, light flavour jet. In order to match the topology, each selected event is required to have at least four jets with $p_\text{T} > 30$\,GeV. In general, the $p_\text{T}$ threshold is $20$\,GeV for central jets with $|\eta|<2.4$, and $40$\,GeV for jets in the forward detector region ($|\eta|>2.4$). To suppress the background, which is dominated by top quark pair production, the Combined Secondary Vertex b tagging algorithm is used to identify b jets. Two independent signal regions (3T and 4T) are defined by requiring either three or four b~tagged jets, plus one untagged jet in every region.

Missing transverse energy is expected to stem from the escaping neutrino. A requirement of $E_{\text{T}}^{\text{miss}}> 35$\,GeV in the muon channel and $E_{\text{T}}^{\text{miss}} > 45$\,GeV in the electron channel is imposed, which drastically reduces the amount of contamination by QCD multi-jet background events.

Roughly 13 signal events are expected in the 3T region after applying the selection. The signal-over-background ratio corresponds to $\sim 0.7\,\%$. The 4T region is purer in the signal ($S/B\sim2.1\,\%$), but has lower yields with only 1.4 expected signal events.

\subsection{Event reconstruction}

Even after a dedicated event selection as described above, the remaining events are dominated by $\text{t}\bar{\text{t}}$ production. Multivariate techniques are therefore needed to further separate the signal process from background, based on the features of tHq and $\text{t}\bar{\text{t}}$ events.

Prior to this, a correspondence between reconstructed jets and the final state objects must be build in order to define the input variables in the most efficient way. For this purpose, two variants for a multivariate jet assignment are used, respectively under signal and $\text{t}\bar{\text{t}}$ hypotheses. Under both hypotheses, the metrics used to assess the correctness of the assignment in simulated events is based on angular matching between jets and partons.

For the jet assignment in a simulated signal event, all possible ways to assign four reconstructed jets to the four final state quarks from $\text{tHq}\to3{\text{bql}\nu}$ are considered, where a correct event interpration is present in the case where four jets can be matched to the appropriate quarks within a cone of size $\Delta R=\sqrt{\Delta \eta^2 + \Delta \phi^2}=0.3$. If the distance between at least one quark and its assigned jet is larger than this threshold, the event interpretation is considered wrong. 
The total number of possible interpretations is reduced by additional requirements: because of b~tagging considerations, b~quarks must be associated to central jets ($|\eta|<2.4$), while only a jet failing the b~tagging requirement can be assigned to the light recoil quark. 

A multivariate analysis tool (MVA) is then trained to distinguish between correct and wrong interpretations with input variables that employ kinematical characteristics of the signal process and include also detector-related information like b~tagging.

The finally chosen event interpretation of any event (from data or simulation) used in the analysis is the one which gives the largest MVA response from all possible tHq jet assignments.

Another MVA is evaluated at the same analysis level for interpretation of events under the assumption that they are semileptonic $\text{t}\bar{\text{t}}$ ones. For this, it has been trained with correct and wrong quark-jet assignments in simulated semileptonic $\text{t}\bar{\text{t}}$ events with $\text{t}\bar{\text{t}}\to2\text{b}2\text{ql}\nu$ in a way analogue to the tHq jet assignment described above. The restriction of combinatorics happens by requiring that only b~tagged jets can be assigned to the two b~quarks. The list of input variables employs kinematics and b~tagging information capturing the characteristics of a semileptonic $\text{t}\bar{\text{t}}$ event.

For application of the training results to all events used in the analysis, the jet assignment yielding the largest MVA response is picked as event interpretation under the $\text{t}\bar{\text{t}}$ hypothesis.

\subsection{Event classification}
\label{sec:classification_bbar}

\begin{figure}[tpb]
\centering
\includegraphics[width=17.55pc]{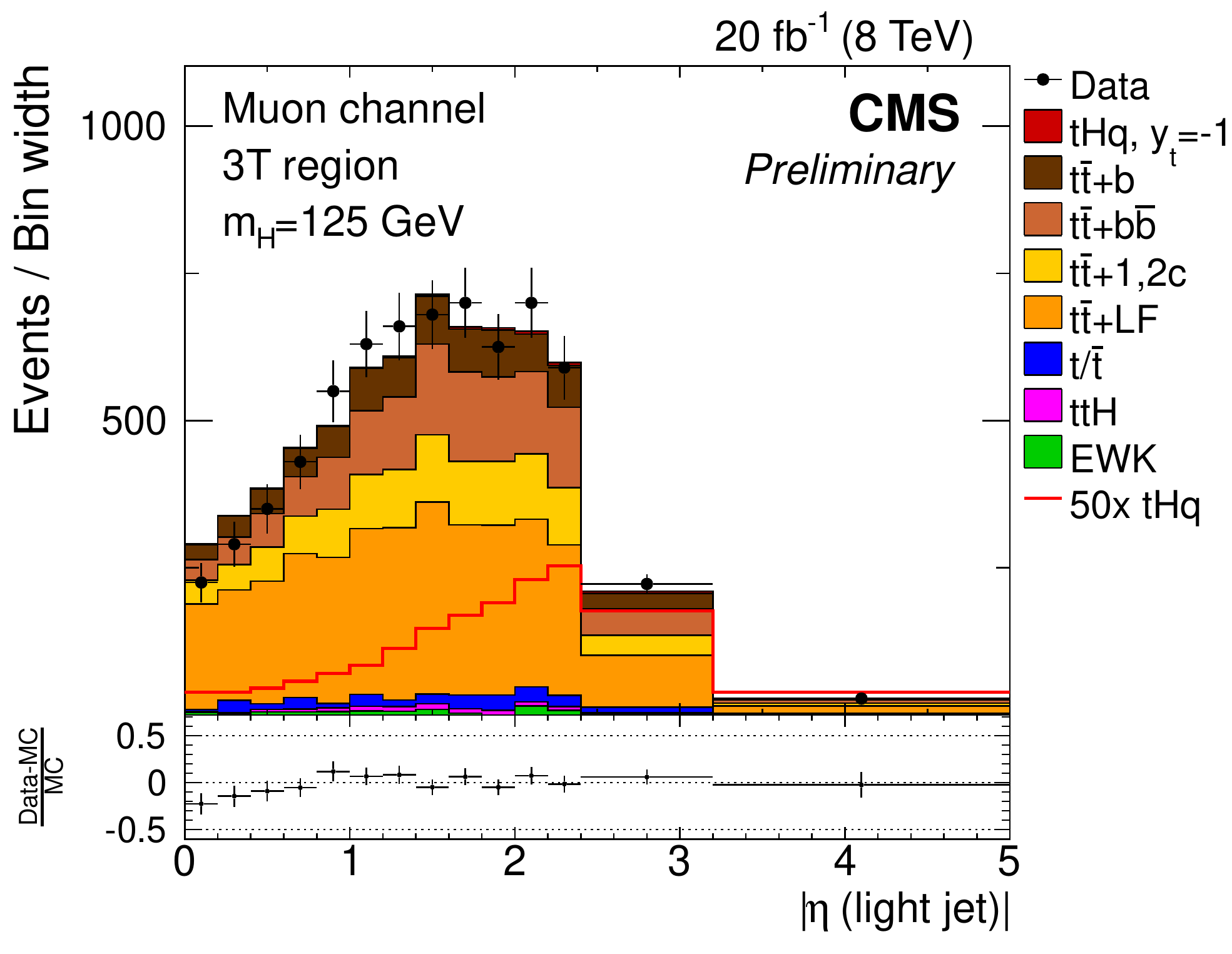}
\hspace{0.3cm}
\includegraphics[width=17.55pc]{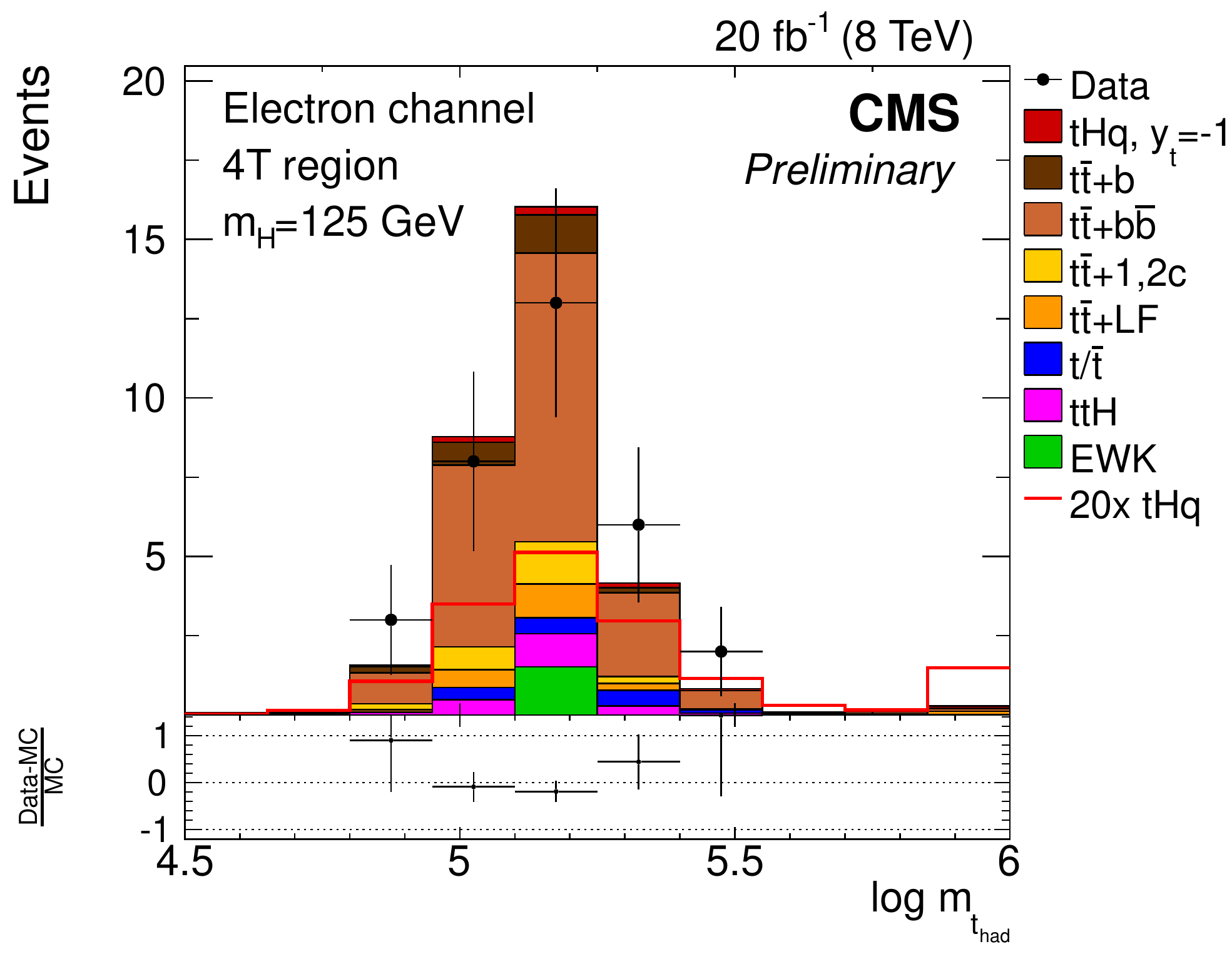}\\
\caption{Two input variables for the final classification MVA. The absolute $\eta$ value of the light jet defined under the tHq jet assignment (left) and the logarithmic mass of the hadronically decaying top quark defined under the semileptonic $\text{t}\bar{\text{t}}$ jet assignment (right). Simulation is normalized to data.}
\label{fig:variables}
\end{figure}

The two types of reconstructions described above are carried out in every event passing the event selection; this allows to construct two sets of observables, where one set presents intrinsic properties of a signal, tHq, event, and the other set consists of variables characteristic for semileptonic $\text{t}\bar{\text{t}}$ events. Together with the lepton charge, which as a global variable is independent from any kind of reconstruction, they form the list of input variables for the final MVA which classifies events as signal- or background-like. Figure~\ref{fig:variables} shows two input variables which provide a high discrimination power between signal and background. Like these two, all used variables show a good agreement between data and simulation.

\section{Results}

\begin{figure}[t]

\begin{minipage}{0.48\textwidth}
\includegraphics[width=17.55pc]{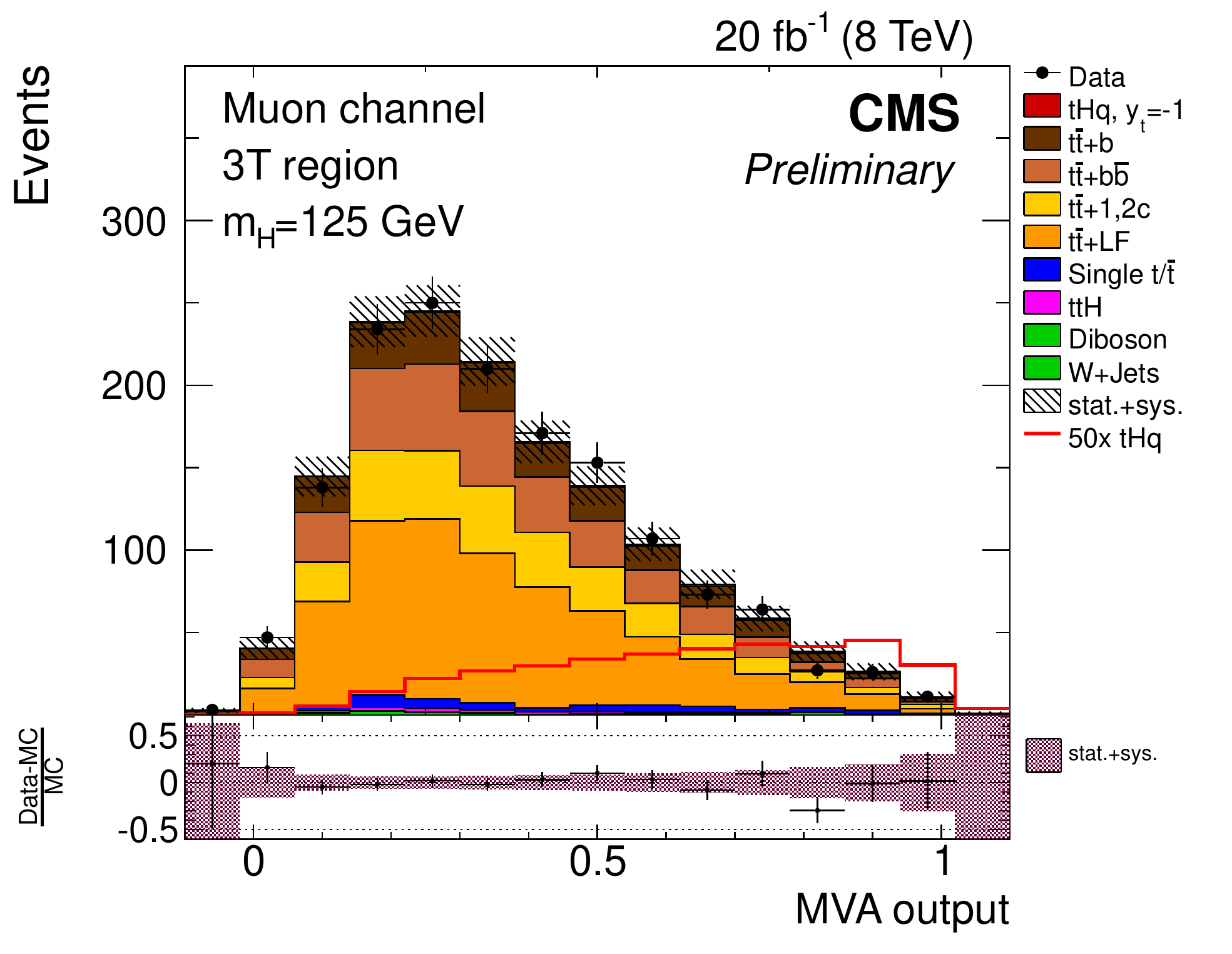}
\caption{\label{fig:MVAoutput}Post-fit MVA output distribution in the 3T muon region as used in the upper $\text{CL}_{\text{S}}$ limit extraction. The signal is scaled by a factor 50 in order to make it visible.}
\end{minipage}\hspace{0.04\textwidth}%
\begin{minipage}{0.48\textwidth}
\vspace{0.43cm}
\hspace{0.5cm}\includegraphics[width=14pc]{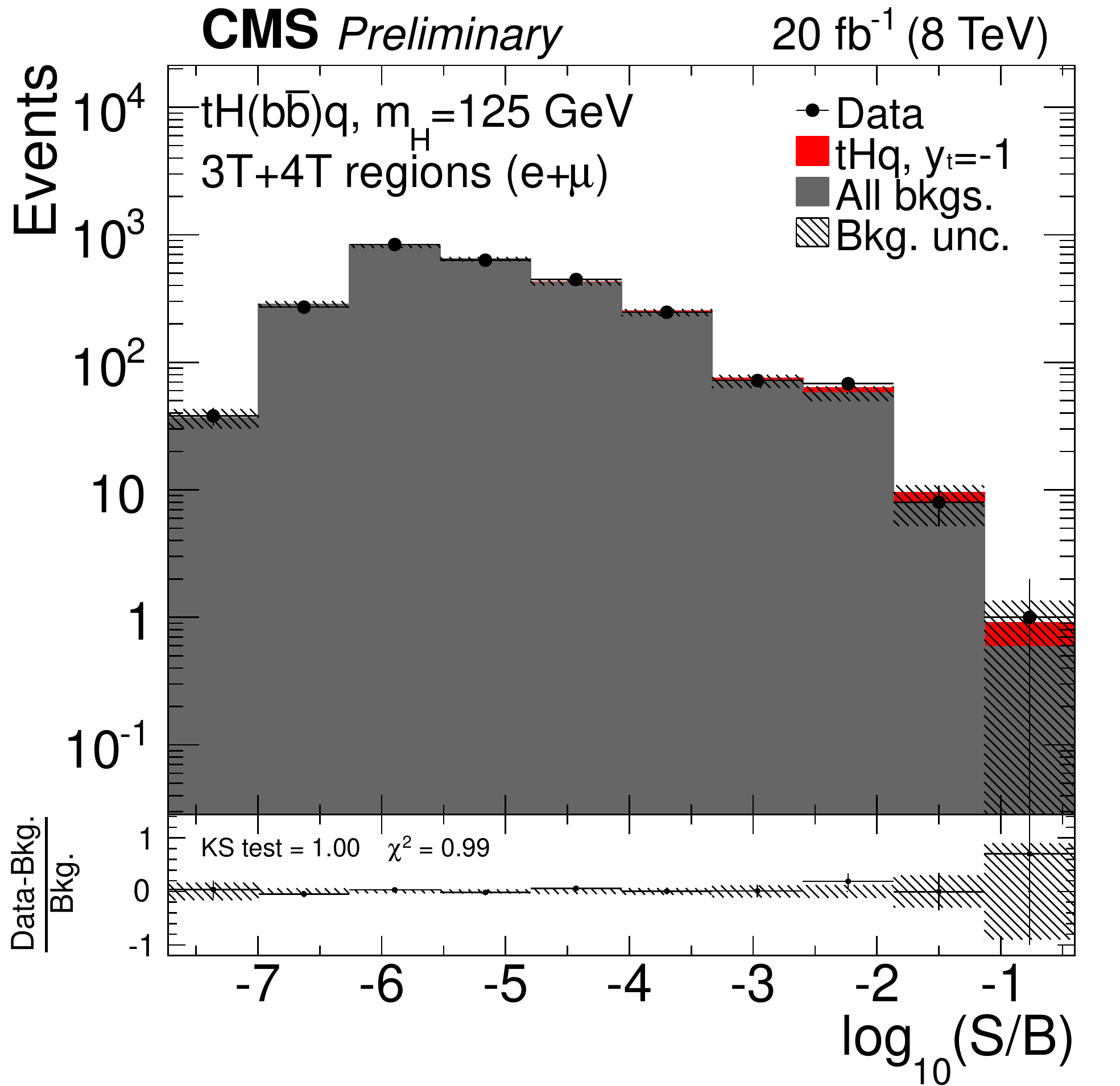}
\caption{\label{fig:SoverB}Events from all regions are sorted according to the $S/B$ ratio in the MVA output distributions. Good agreement of the data with the background-only hypothesis is observed.}
\end{minipage} 
\end{figure}

The upper limit on the tHq production cross section is obtained by simultaneously fitting the classification MVA output in the 3T and 4T analysis regions, for both muon and electron channel. Figure~\ref{fig:MVAoutput} exemplarily shows the post-fit MVA output distribution in the 3T region in the muon channel. The observed upper limit on $\sigma_{95\,\%}/\sigma_{\text{tHq}}$ with $y_\text{t}=-1$ calculated from these distributions with a fully frequentist approach is found to be 7.57, with an expected search sensitivity of $5.14$. Accordingly, the analysis is able to exclude tHq production with $y_\text{t}=-1$ below $\sigma_{\text{tHq}}=1.77$\,pb.

A cross-check approach which uses a data-driven technique to model the dominant $\text{t}\bar{\text{t}}$ background in the signal regions by employing b tagging and mistagging efficiences in a 2T control region gives consistent results. All 95\%~confidence level exclusion limits are summarized in Table~\ref{tab:results}.

\renewcommand{\arraystretch}{1.5} 
 \begin{table}[!ht]
  \caption{Expected and observed 95\%~confidence level upper limits on $\sigma/\sigma_{y_\text{t} = -1}$ for the combination of all analysis bins (3T+4T) for the MC-driven and the data-driven approach. For the expected limits also the $+1\,\sigma$ and $-1\,\sigma$ deviations are displayed.}
 \begin{center}
 \scalebox{0.88}{
 \begin{tabular}{ccc} \hline\hline
         & Expected & Observed       \\ \hline
 \textbf{MC-driven} & $\mathbf{5.14^{+2.14}_{-1.44}}$ & $\mathbf{7.57}$ \\
 Data-driven cross-check & $6.24^{+2.26}_{-1.71}$  & 6.95 \\
 \hline \hline
 \end{tabular}
 }
 \end{center}

 \label{tab:results}
 \end{table}

\end{document}